\documentclass[prl,twocolumn,preprintnumbers,amsmath,amssymb]{revtex4}

\usepackage{times}
\usepackage{natbib}
\usepackage{amssymb,amsbsy,amsmath,amsfonts}
\usepackage{graphicx}
\usepackage{epsf,epsfig,float,latexsym,amsthm,fancyhdr,rotating}
\usepackage{graphics,psfrag,longtable}

\usepackage[svgnames]{xcolor}

\usepackage{ulem}

\begin{document}
\preprint{UCSD/PTH 14-07}

\title{$SU(2)\times U(1)$ gauge invariance and the shape of new physics in  rare $B$ decays}

\author{R.~Alonso$^{1}$}
\author{B.~Grinstein$^{1}$}
\author{J. Martin Camalich$^{1,2}$}

\affiliation{
$^1$Dept. Physics, University of California, San Diego, 9500 Gilman Drive, 
La Jolla, CA 92093-0319, USA\\
$^2$PRISMA Cluster of Excellence Institut f\"ur Kernphysik, 
Johannes Gutenberg-Universit\"at Mainz, 55128 Mainz, Germany}

\begin{abstract}

New physics effects in $B$ decays are routinely
modeled through operators invariant under the strong and
electromagnetic  gauge symmetries. Assuming the scale for new physics is well above the electro-weak
scale, we further require invariance under the full Standard-Model
gauge symmetry group. Retaining up to  dimension-6 operators, we unveil new constraints between different 
new-physics operators that are assumed to be independent in
the standard phenomenological analyses. We illustrate this approach by analyzing the constraints on new physics
 from rare $B_{q}$ (semi-)leptonic decays.
\end{abstract}

\maketitle

\paragraph{Introduction.}
The exploration of the energy regime of electro-weak symmetry breaking (EWSB) at the
Large Hadron Collider (LHC) has unveiled a scalar boson~\cite{Aad:2012tfa,Chatrchyan:2012ufa} resembling the Standard Model (SM) 
singlet component of the Higgs doublet and no other particle. 
If one therefore assumes, as the experimental evidence suggests, that the scale of new physics (NP), $\Lambda$,
is above the EWSB scale an effective field theory (EFT) built exclusively from SM fields can be used. 
In this widespread and fruitful scheme, higher-dimension operators  suppressed
by powers of the NP scale encode deviations of the SM in a
generic and model-independent manner~\cite{Buchmuller:1985jz,Grzadkowski:2010es}. 
The only requirements imposed on the operators are Lorentz 
and $SU(3)_c\times SU(2)_L\times U(1)_Y$ gauge symmetries.

These simple assumptions lead, as this letter is meant to show, to
phenomenological consequences not only for physics at the EWSB scale
but also for physics well below such scale. Furthermore, the
consequences not only affect the ``size'' of the contribution of new
physics to low energy processes, but also the ``shape'' or correlation
among different operators. These extra constraints in the low energy
Lagrangian are due simply to $SU(3)_c\times SU(2)_L\times U(1)_Y$
  invariance.  More specifically, there are three important ways in
  which the low energy EFT is further
  constrained:\\
{\it (i)} The operators must originate in those of an EFT with explicit electroweak symmetry;\\
{\it (ii)}  The coefficients of operators are not all independent, as they 
may be related by their origin in the underlying spontaneously broken electroweak group; and\\
{\it (iii)} Some of the
coefficients of the low energy EFT may be
constrained by seemingly unrelated high energy processes.\\
 The later occurs,
for example, when the low energy operator arises from integrating out
a heavy field, like the Higgs, from an operator which itself produces
effects observable in the decay of the heavy field.

To illustrate the aforementioned effects we consider rare, flavor
changing-neutral (FCN) $B$-meson semi- or purely-leptonic decays, where,
to our knowledge, such an analysis has not yet been carried out
fully. We shall distinguish between three different scales:
$i$)~the NP scale $\Lambda$, $ii$)~the EWSB scale, $\langle
H^\dagger H \rangle= v^2/2$, and $iii$)~the low scale, $\mu$,
in this case of the order of the bottom quark mass. We assume
the following hierarchy of mass scales $\mu \ll v\ll\Lambda$.
The reduced set of observables that will be studied here
allow us to focus on a subgroup of operators rather than the most general EFT consistent with
electroweak symmetry, which is left for future work.


\paragraph{Low energy: B-meson semi-leptonic Lagrangian.}
At energies around the bottom quark mass, the EFT  Lagrangian is built from 
the light  fields: the SM particle content except the $W$ 
and $Z$ bosons, the Higgs boson, and the top quark. In addition
the EFT  Lagrangian respects the gauge symmetries manifest at this scale, 
namely $SU(3)_c\times U(1)_{em}$. As we will show,
not all of the possible operators constructed in this way are 
compatible with an effective Lagrangian invariant under $SU(2)_L\times U(1)_Y$. 

To leading order in $G_F=1/(\sqrt2 v^2)$ the effective Lagrangian for $\Delta B=1$
processes is~\cite{Grinstein:1988me,Buchalla:1995vs,Chetyrkin:1996vx}
\begin{equation}
\label{eq:Heff}
  {\cal{L}}_{\rm eff}=-
  \frac{4 G_F}{\sqrt{2}}\sum_{p=u,c}\lambda_{ps}\left(C_1 \mathcal{O}_1^p+C_2\mathcal{O}_2^p +\sum_{i=3}^{10} C_i  \mathcal{O}_i\right),
\end{equation}
with $\lambda_{ps}=V_{pb}^{} V_{p s}^\ast$. The ``current-current'' operators,
$\mathcal{O}_{1,2}^p$, ``QCD-penguins,'' $\mathcal{O}_{3,\ldots,6}$, and
``chromo-magnetic operator,'' $\mathcal{O}_{8}$, do not contribute to
$B_s\to\ell\bar\ell$ and their contribution to $B\rightarrow K^{(*)}\ell\bar\ell$
requires an electromagnetic interaction (they contribute to the 
``non-factorizable'' corrections, in the language of QCD factorization~\cite{Beneke:2001at}). We therefore focus on the electromagnetic penguin,
$\mathcal{O}_7$ and the semileptonic operators, $\mathcal{O}_{9,10}$, defined as
\begin{align}
  \mathcal{O}_7 & = 
    \frac{e}{(4 \pi)^2} \overline m_b [\bar{s} \sigma^{\mu\nu}P_R\, b] F_{\mu\nu}, \label{eq:SMbasis} \\
  \mathcal{O}_9 & = 
    \frac{e^2}{(4 \pi)^2} [\bar{s} \gamma_\mu  P_Lb][\bar{l} \gamma^\mu l], ~
  \mathcal{O}_{10}  = 
    \frac{e^2}{(4 \pi)^2} [\bar{s} \gamma_\mu  P_L b][\bar{l} \gamma^\mu \gamma_5 l],  \nonumber
\end{align}
where $b,s,l$ stand for the bottom and strange quarks and a charged lepton, respectively, $F_{\mu\nu}$ is the photon field strength and
$P_{R,L}=(1\pm\gamma_5)/2$.

In addition, BSM physics can generate chirally-flipped ($b_{L(R)}\rightarrow b_{R(L)}$)
versions of these operators, $\mathcal{O}_{7,\ldots, 10}'$, and also four scalar
and two tensor operators~\cite{Bobeth:2007dw},
\begin{align}
  \mathcal{O}_S^{(\prime)} & = \frac{e^2}{(4 \pi)^2} [\bar{s} P_{R(L)} b] [\bar{l} l], ~
  \mathcal{O}_P^{(\prime)}  = \frac{e^2}{(4 \pi)^2} [\bar{s} P_{R(L)} b] [\bar{l} \gamma_5 l],
\label{eq:nonSM:scalars} \\
  \mathcal{O}_T   & = \frac{e^2}{(4 \pi)^2} [\bar{s} \sigma_{\mu\nu} b][\bar{l} \sigma^{\mu\nu} l], ~
  \mathcal{O}_{T5}  = \frac{e^2}{(4 \pi)^2} [\bar{s} \sigma_{\mu\nu} b][\bar{l} \sigma^{\mu\nu} \gamma_5 l].
  \label{eq:nonSM:tensors}
\end{align}
Note that there are only two possible non-vanishing tensor
  operators ~\footnote{$(\bar l \sigma^{\mu\nu} P_R l)
      (\bar s\sigma_{\mu\nu} P_L b)=(\bar l \sigma^{\mu\nu} P_L
      l) (\bar s\sigma_{\mu\nu} P_R b)=0$ identically.}.
These, together with those in Eq.~(\ref{eq:SMbasis}) and their
chirally-flipped counterparts, constitute the most
general basis for the Lagrangian describing $B_s$ (semi-)leptonic rare decays.   
 In this construction, the 12 coefficients in the
EFT Lagrangian of these 12 distinct operators are {\it a priori}
independent.  However, as discussed in the next section,
if the NP lies above the EW scale the number of free coefficients is reduced
to~8.

The same effective Lagrangian (and following discussions) can be
applied to $b\rightarrow d$ decays by replacing a $d$-quark for
the $s$-quark in Eq.~(\ref{eq:Heff}), and accounting for the
difference in CKM elements by replacing $\lambda_{pd}$ for
$\lambda_{ps}$ throughout. The Wilson coefficients are not necessarily
the same as in the $b\rightarrow d$ transitions, and comparing the 
two sets of coefficients would give
information about flavor violation of presumed NP.
Similarly, Wilson coefficients could also depend on the family of the
leptons, which would result in  lepton universality violation.

\paragraph{High energy: New Physics above the EWSB scale.}
If the operators appearing in the effective Lagrangian are generated by
physics at a scale $\Lambda$  above the EWSB scale, $v\ll\Lambda$ they 
must originate from operators manifestly $SU(3)_c\times SU(2)_L\times U(1)_Y$ invariant. 
The fields at our disposal for the construction of such Lagrangian are the chiral fermions
$q_L=(u_L, d_L)^T,\,\ell_L=(\nu_L, l_L)^T,\,u_R$, $d_R\,,e_R,$ the Higgs doublet $H$ and
covariant derivatives containing gluons, weak-isospin and hypercharge vector bosons.
We will work in the basis in which the down-type Yukawa matrix is diagonal and write
the quark doublets as $q_{d}= (u_{jL }V^*_{jd},d_{L})$, $q_{s}= (u_{jL }V^*_{js},s_{L})$ and
$q_{b}= (u_{jL }V^*_{jb},b_{L})$.

We restrict attention to BSM operators of dimension 6. The
effective Lagrangian takes the form $\mathcal{L}_{BSM}=\tfrac1{\Lambda^2}\sum_i C_i
Q_i$. The relevant operators for the 
study of rare (semi-)leptonic decays in the $B_q$ system are either dipole-like,
\begin{align}
Q_{dW}         & =g_2(\bar q_s \sigma^{\mu\nu} b_R) \tau^I H\, W_{\mu\nu}^I,~
Q_{dB}        =g_1(\bar q_s \sigma^{\mu\nu} b_R) HB_{\mu\nu},\nonumber\\
Q_{dW}'        & =g_2H^\dagger\tau^I(\bar s_R \sigma^{\mu\nu} q_b)  W_{\mu\nu}^I,~
Q_{dB}'        =g_1H^\dagger(\bar s_R \sigma^{\mu\nu} q_b)B_{\mu\nu}, 
  \label{eq:L6typeI}
\end{align}
Higgs-current times fermion-current,
\begin{align}\nonumber
Q^{(1)}_{H q} &=\left(H^\dagger\,i\overleftrightarrow{D}_\mu H \right)(\bar q_s \gamma^\mu q_b) \\ 
Q^{(3)}_{H q} &=H^\dagger\,i(\tau^I\overrightarrow{D}_\mu- \overleftarrow{D}_\mu\tau^I )H (\bar q_s \tau^I\gamma^\mu q_b)\label{eq:L6typeII}\\
Q_{H d} &=\left(H^\dagger\,i\overleftrightarrow{D}_\mu H \right)(\bar s_R \gamma^\mu b_R)\nonumber
\end{align}
or four-fermion,
\begin{align}
Q_{\ell q}^{(1)}                & =(\bar \ell \gamma_\mu \ell)(\bar q_s \gamma^\mu q_b), &
Q_{\ell q}^{(3)}                &= (\bar \ell \gamma_\mu \tau^I \ell)(\bar q_s \gamma^\mu \tau^I q_b), \nonumber\\
Q_{ed}                      & =(\bar l_R \gamma_\mu l_R)(\bar s \gamma^\mu b_R), &
Q_{\ell d}               & =(\bar \ell \gamma_\mu \ell)(\bar s \gamma^\mu b_R), \nonumber\\ 
Q_{qe}               & =(\bar q_s \gamma_\mu q_b)(\bar l \gamma^\mu l_R), &
Q_{\ell edq} &= (\bar q_s b_R)(\bar   l_R \ell), \nonumber \\
Q_{\ell edq}' &=(\bar \ell\,  l_R) (\bar s_R q_b), & &\label{eq:HE4f}
\end{align}
where color and weak-isospin indices are omitted and $\tau^I$ stand for the Pauli matrices in $SU(2)$-space.
Primed operators correspond to a different flavor entry of the hermitian conjugate of the unprimed operator.

This Lagrangian cannot be compared still with that of Eq.~(\ref{eq:Heff}); one has to
integrate out the heavy degrees of freedom, {\it i.e.}, $Z$, $W$,
$t$ and $H$, and run it down to $\mu_b$.  The first step yields
four-fermion and dipole operators as in Eqs.~(\ref{eq:SMbasis},
\ref{eq:nonSM:scalars}). Remarkably,  no new tensor-like operators \eqref{eq:nonSM:tensors} 
appear after integration of $W$ and $Z$ bosons at leading order.
By contrast, new contributions to the coefficients of 
$\mathcal{O}^{(\prime)}_{9,10}$ are indeed
generated by the operators in Eq.~(\ref{eq:L6typeII}).

Explicitly, the connection with the Lagrangian of Eq.~(\ref{eq:Heff}), 
at the scale $M_W$ is~\footnote{For a complete connection, the
coefficients should be previously run from $\Lambda$ to $M_W$~(\cite{Jenkins:2013zja,Jenkins:2013wua,Alonso:2013hga}),
this is taken into account for the numerical bounds on $\Lambda$.}, 
for scalar and tensor type operators:
\begin{eqnarray}
&&C_{S}^{l}=-C_{P}^{l}=\frac{4\pi^2}{e^2\lambda_{ts}}\frac{v^2}{\Lambda^2}C_{\ell edq},\nonumber\\
&&C_{S}^{l\prime}=C_{P}^{l\prime}=\frac{4\pi^2}{e^2\lambda_{ts}}\frac{v^2}{\Lambda^2}C_{\ell edq}',\nonumber\\
&&C_{T}=C_{T5}=0,\label{eq:SMconstraintsI}
\end{eqnarray}
for dipole operators:
\begin{eqnarray}
&&C_7^{(\prime)}=\frac{8\pi^2}{ y_b \lambda_{ts}}\frac{v^2}{\Lambda^2}\left(C_{dB}^{(\prime)}-C_{dW}^{(\prime)} \right),\nonumber 
\label{eq:SMconstraintsII}
\end{eqnarray}
and for the current-current type of leptonic operators:
\begin{align}
C_9=&\frac{4\pi^2}{e^2\lambda_{ts}}\frac{v^2}{\Lambda^2}\left(C_{qe}+C_{\ell q}^{(1)}+C_{\ell q}^{(3)}-(1-4s_W^2)(C_{Hq}^{(1)}+C_{Hq}^{(3)}) \right),\nonumber\\ 
C_{10}=&\frac{4\pi^2}{e^2\lambda_{ts}}\frac{v^2}{\Lambda^2}\left(C_{qe}-C_{\ell q}^{(1)}-C_{\ell q}^{(3)}+(C_{Hq}^{(1)}+C_{Hq}^{(3)}) \right),\nonumber\\
C_9'=&\frac{4\pi^2}{e^2\lambda_{ts}}\frac{v^2}{\Lambda^2}\left(C_{ed}+C_{\ell d} -(1-4s_W^2)C_{Hd} \right),\nonumber\\
C_{10}'=&\frac{4\pi^2}{e^2\lambda_{ts}}\frac{v^2}{\Lambda^2}\left(C_{ed}-C_{\ell d} +C_{Hd}\right).  \nonumber 
\end{align}
Equation~\eqref{eq:SMconstraintsI} shows explicitly what has been advertised in the introduction:\\
{\it (i)}  Some operators cannot be generated in the EFT ($C_T=C_{T5}=0$).\\
{\it (ii)} There are correlations between nonvanishing coefficients
($C_S=-C_P$ and $C_S'=C_P'$). \\
{\it (iii)} The contributions to some EFT coefficients may be 
subject to constraints arising purely from high energies ({\it e.g.},
$Q_{dW}^{(\prime)}$,  $Q_{dB}^{(\prime)}$, $Q_{Hq}$ and $Q_{Hd}$  contribute to
flavor-violating $Z$ and $H$  decays).

The reduction in the number of structures occurs only for scalar, pseudo-scalar and tensor operators. 
The reason for this reduction is invariance under hypercharge: the tensor-like operators
simply cannot be promoted to be $U(1)_Y$ invariant, and for scalar and pseudo-scalar $U(1)_Y$ requires
the leptons to have definite chirality dependent on the $b$ quark chirality. 
For the rest of operators the coefficients are independent linear combinations. However, note that there are additional 
correlations between the neutral current and the charged current version of the operators
that arise from operators involving doublets. While these play no role directly in FCN
 leptonic decays of $B$ mesons, they may 
give rise to additional constraints on the effects of NP.

Violations to the relations of
  Eq.~(\ref{eq:SMconstraintsI})  of order $v^2/\Lambda^2$ arise from  dimension-8 operators like $\bar q H b_R\,
\bar \ell H l_R$ and possibly of
  order $g_{EW}^2/16\pi^2$ from 1-loop matching.

\paragraph{Consequences in $B^0_{s,d}\rightarrow l^+l^-$.}
 A powerful  probe of NP is the decay $B^0_{s,d}\rightarrow l^+l^-$. 
In the SM it is first induced  at 1-loop level and  
is chirally suppressed. Moreover, the hadronic matrix element 
is determined fully by $B^0_{s,d}$ decay constants $F_{B_{s,d}}$, which
are calculated in lattice QCD~\cite{Aoki:2013ldr}. 

The SM predictions for the branching fractions, $\overline{\mathcal{B}}$,  have been worked out
to high accuracy. For the muonic and electronic modes they 
currently are~\cite{Bobeth:2013uxa}:
\begin{align}
\overline{\mathcal{B}}_{s\mu}=& 3.65(23)\times10^{-9}, 
&\overline{\mathcal{B}}_{d\mu}=& 1.06(9)\times 10^{-10}, \nonumber\\
\overline{\mathcal{B}}_{se}=& 8.54(55)\times10^{-14}, 
&\overline{\mathcal{B}}_{de}=& 2.48(21)\times10^{-15}, \label{eq:Bsllth}
\end{align}
where the overline indicates  untagged, time-integrated rates (as
  required by the sizable 
width difference in the $\bar{B}_s-B_s$ system, although not  for $B_d$~\cite{DeBruyn:2012wk}). 

The muonic modes have been recently measured by LHCb~\cite{Aaij:2012nna,Aaij:2013aka}
and CMS~\cite{Chatrchyan:2013bka}, and an average of the results leads 
to~\cite{LHCbCMScombo}:
\begin{align}
&\overline{\mathcal{B}}_{s\mu}^{\rm expt}=2.9(7)\times10^{-9}, 
&\overline{\mathcal{B}}_{d\mu}^{\rm expt}=3.6^{+1.6}_{-1.4}\times10^{-10}, \label{eq:Bsmumuexpt}
\end{align}
where the $\overline{\mathcal{B}}_{d\mu}$ mode is not statiscally significant yet ($<3\sigma$).  
For the electronic modes we currently have only upper bounds at 95\% C.L.~\cite{HFAG}:
\begin{align}
&\overline{\mathcal{B}}_{se}^{\rm expt}<2.8\times10^{-7},  
&\overline{\mathcal{B}}_{de}^{\rm expt}<8.3\times10^{-8}. \label{eq:Bseeexpt}
\end{align}

Useful quantities to compare the theory to are the ratios~\cite{DeBruyn:2012wk}:
\begin{equation}
\overline{R}_{ql}=\frac{\overline{\mathcal{B}}_{ql}}{\left(\overline{\mathcal{B}}_{ql}\right)_{\rm SM}}=
\frac{1+\mathcal{A}^{ll}_{\Delta \Gamma}\,y_q}{1+y_q}\left(|S|^2+|P|^2\right), \label{eq:R}
\end{equation}
where $y_q=\tau_{B_q}\Delta\Gamma_q/2$, $\mathcal{A}^{ll}_{\Delta \Gamma}$ is the 
mass eigenstate rate asymmetry~\cite{DeBruyn:2012wk} and: 
\begin{gather}
S=\sqrt{1-\frac{4m_l^2}{m_{B_q}^2}}\frac{C_S-C_S'}{r_{ql}},~~
P=\frac{C_{10}-C_{10}'}{C_{10}^{\rm SM}}+\frac{C_P-C_P'}{r_{ql}},\nonumber \\ 
\text{where}~~~~~~~~ r_{ql}=\frac{2m_l\,(m_b+m_q)C_{10}^{SM}}{m_{B_q}^2}.
\label{eq:SandP}
\end{gather}

The contributions of $C_S^{(\prime)}$ and $C_P^{(\prime)}$ are
enhanced by the factor ${m_{B}}/{m_l}$, so below we will
neglect the NP in $C_{10}^{(\prime)}$  for simplicity. 
The decay 
rate is only sensitive to the differences $(C_P-C_P')$ and $(C_S-C_S')$ 
so the  sums,  $(C_P+C_P')$ and
    $(C_S+C_S')$, need to
  be constrained through other  means.


\begin{figure}[h]
\centering
      \includegraphics[scale=0.15]{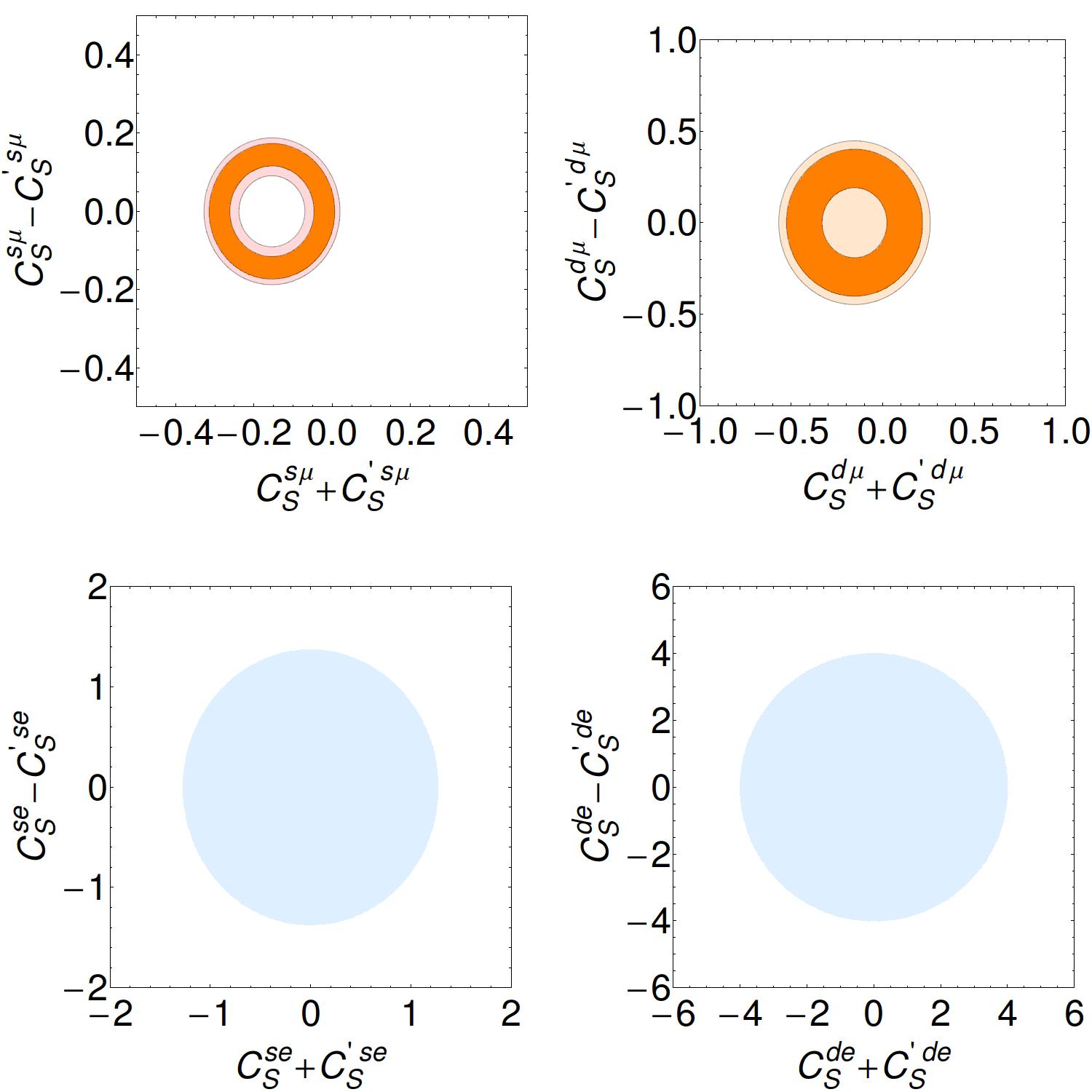}
\caption{\label{fig:Bqll} In the upper panel we show the limits at 1$\sigma$
  and 3$\sigma$
on the scalar Wilson coefficients that are induced by the 
experimental $\overline{\mathcal{B}}_{ql}$ in Eq.~(\ref{eq:Bsmumuexpt}),
 where the corrections by mixing have been taken into account. 
For the electronic modes in the lower panel, we only show the 3$\sigma$ allowed
regions~(\ref{eq:Bseeexpt}). In both cases the Wilson coefficients are understood to be 
renormalized at $\mu=m_b$. }
\end{figure}

 Introducing
the hypothesis of this work, we impose~(\ref{eq:SMconstraintsI}) in Eq.~(\ref{eq:R}) 
and~(\ref{eq:SandP}) so that now
\begin{equation}
\overline{R}_{ql}\simeq\frac{|C_S-C_S'|^2}{r^2_{ql}}+\left|1-\frac{C_S+C_S'}{r_{ql}}\right|^2, \label{eq:Rewconst}
\end{equation}
where we have neglected $y_s=0.075(12)\%$~\cite{Aaij:2013oba} and the phase space factor for clarity.
 In addition to the reduction of free parameters from 4 to 2 in the
scalar and pseudo-scalar sector, now these two parameters enter
the decay rate in two orthogonal linear combinations. As a result the
$B_q\rightarrow l^+l^-$ branching fraction alone bounds {\it all
directions} in our two parameter space.
In particular, for real Wilson coefficients, the bound  of Eq.~(\ref{eq:Rewconst})
defines a circle in parameter space centered at
$(C_S+C_S',C_S-C_S')=(r_{ql},0)$ with radius $|r_{ql}|\sqrt{\overline{R}_{ql}^{\rm expt}}$.

The contour plots in Fig.~\ref{fig:Bqll} show these circular 
 bounds with the radius of the circle in the muonic cases determined 
by $|r_{q\mu}|\simeq0.16$. 
This shape is in contrast with the  bands, experimentally unconstrained in one direction,  that would be obtained in the standard analysis.
Note that improving the experimental accuracy 
in these modes will only reduce the width of the ring and that breaking 
the degeneracy will require other observables. One attractive possibility 
 is the observable $\mathcal{A}^{\mu\mu}_{\Delta \Gamma}$, which 
 may be obtained by measuring the effective $B_s\rightarrow\mu^+\mu^-$ 
lifetime~\cite{DeBruyn:2012wk}.

 For the electronic modes, 
$|r_{qe}|\sim10^{-3}$ and the strength of the limits
in the parameter space  is governed by the size
of $\overline{R}_{qe}^{\rm expt}$. Hence, improved experimental 
bounds on the branching fractions have the potential to probe 
extremely high energies through the scalar operators.


To quantify this, the bounds on the scale $\Lambda$ 
for the different decay rates in Eq.~(\ref{eq:Bsllth}) can be computed making
use of Eq.~(\ref{eq:SMconstraintsI}) and assuming naturalness, namely $C_{ledq}^{(\prime)}(\Lambda)\simeq1$. 
To compute these bounds the running is taken into account in the two
intervals: {\it i)} from $m_b$ to $\sim$$M_W$ where
 QCD at 1 loop dominates~\cite{Bobeth:2007dw} and strengthens the bound by $\sim0.74$\,,
{\it ii)} from $~M_W$ to $\Lambda$ 
using the recently-computed anomalous dimension~\cite{Jenkins:2013zja,Jenkins:2013wua,Alonso:2013hga}, which 
gives an extra factor $\sim0.66$. The bounds from $\bar B_{s\mu}$, $\bar B_{d\mu}$, $\bar B_{se}$, and $\bar B_{de}$, are respectively $78$\,TeV, $130$\,TeV, $36$\,TeV, and $49$\,TeV.
Note that $B_d$ mesons decays do better at constraining new physics due to the CKM suppressed SM background. 
In particular, $\overline{\mathcal{B}}_{de}$ can supersede the present bounds for
a precision of $10^{-10}$. 

\paragraph{Consequences in $B\rightarrow K^{(*)}l^+l^-$.}
The exclusive semileptonic $B$ decays are also powerful 
flavor laboratories. As 3- and 4-body decays, their angular 
distributions lead to a rich and non-trivial phenomenology 
which could potentially unveil NP surfacing through 
various operators (see e.g. Refs.~\cite{Bobeth:2007dw,Altmannshofer:2008dz,
Bobeth:2013uxa,Jager:2012uw,Descotes-Genon:2013vna} and 
references therein).

The effects of scalar or tensor operators in 
 $B\rightarrow K^{(*)}l^+l^-$ have been considered by 
several authors~\cite{Bobeth:2007dw,Altmannshofer:2008dz,Bobeth:2013uxa}.
The first immediate consequence of our analysis in these 
decays is that tensor operators can be ignored altogether (up to 
$\mathcal{O}(v^2/\Lambda^2)$ corrections). This observation 
should lead to a considerable simplification in the theoretical 
analyses of the angular observables~\cite{Bobeth:2007dw,Bobeth:2013uxa}.

The scalar operators contribute to the total decay rates 
$B\rightarrow K^{(*)}l^+l^-$, providing another experimental
input to resolve degeneracies. In practice, however, any sensitivity 
is blurred by the SM contribution which depends on quite 
uncertain hadronic form factors~\cite{Jager:2012uw}. As an example, 
the coefficient $I_6^c(q^2)$ in the angular distribution in the $K^{*}$ mode
 is directly proportional to the combination
$|C_S-C_S'|^2$, and it is a null test of the SM~\cite{Altmannshofer:2008dz,Jager:2012uw}
but  the contribution is suppressed by $m_l$ so that the observable is not 
competitive with purely leptonic decays. 

In the case of the $K$ mode the two angular observables $A_{FB}$ 
and $F_H$~\cite{Bobeth:2007dw} are also null tests of the SM
and receive contributions from  $(C_S+C_S')$ and 
$(C_P+C_P')$. In the standard analysis these
observables provide sensitivity to the orthogonal directions scanned 
in $B_q\rightarrow l^+l^-$ and in our case could lift the degeneracies
in Fig.~\ref{fig:Bqll}.
However,   at low $q^2$ the scalars appear suppressed by either $m_l$ 
in $A_{FB}$ or by a kinematical factor in $F_H$~\cite{Bobeth:2007dw}
\footnote{The kinematical suppression in $F_H$ is lifted 
in the low-recoil region~\cite{Bobeth:2013uxa}.}.

%
As a final example let us comment on the impact of our analysis in
lepton universality violation in $B^+\rightarrow K^+l^+l^-$ decays~\cite{Hiller:2003js}.
Recently the LHCb collaboration~\cite{Aaij:2014ora} has reported a deficit in muonic decays with respect to electronic ones
in the $[1,6]$ GeV$^2$ bin
with a significance of $2.6\sigma$:
\begin{equation}
R_K\equiv \frac{\mbox{Br}\left(B^+\rightarrow K^+ \mu\mu \right)}{\mbox{Br}\left(B^+\rightarrow K^+ ee\right)}=0.745^{+0.090}_{-0.074}{\rm (stat)}\pm0.036 {\rm (syst)}. \label{eq:RKexpt}
\end{equation}
In the SM, $R_K$ is given very accurately, $R_K=1.0003(1)$~\cite{Bobeth:2007dw}, 
since the hadronic contributions cancel  in the ratio
 to very good 
approximation. In Ref.~\cite{Bobeth:2007dw}
possible scenarios with sizable scalar operators 
were shown to produce large effects in $R_K$. Our analysis shows that  the bounds from the fully leptonic decay suffice to 
exclude the possibility of scalar operators accounting for~(\ref{eq:RKexpt}), since at 95\% C.L. we have:
\begin{equation}
R_K\in[0.982,1.007].\label{eq:RKscalar}
\end{equation}

In light of this and the absence of tensors, we conclude that 
a large lepton universality violation in $R_K$ could be only produced by the  
operators $\mathcal{O}_9^{(\prime)}$ and  $\mathcal{O}_{10}^{(\prime)}$.
Unfortunately these are not very well bound, especially for the electronic
case, so different scenarios of NP could currently explain~(\ref{eq:RKexpt}).
For example one could entertain the possibility of a sizable and negative 
effect in $C_9$ affecting only the muonic mode, $\delta C_9^{\mu}=-1$. 
In this scenario one obtains $R_K\simeq0.79$. As a side remark,
it is worth emphasizing that such a negative NP contribution
to $\mathcal{O}_9^{(\prime)}$ has been argued to be necessary to 
understand the current $b\rightarrow s\mu\mu$ data set~\cite{Descotes-Genon:2013wba,Altmannshofer:2013foa,Beaujean:2013soa,Horgan:2013pva}.

\paragraph{Conclusions.}
We have discussed a novel approach to the study of new-physics effects
in the $B_{q}$ (semi-)leptonic decays. This relies on the assumption
that the new dynamics enter at a scale $\Lambda\gg v$, and it is based on the 
(tree-level) matching of the effective weak Lagrangian customarily used in 
the phenomenological analyses, $\mathcal{L}_W$, to the most general 6-dimensional 
Lagrangian invariant under the SM gauge group (as done,
  customarily,  in the analysis of other weak
hadronic processes like nuclear and neutron $\beta$-decays~\cite{Bhattacharya:2011qm,Gonzalez-Alonso:2013ura}).

As a direct consequence of $SU(2)_{ L}\times U(1)_{ Y}$ invariance, new constraints
correlate the operators in $\mathcal{L}_W$. For example, in rare $B_q$ (semi-)leptonic decays
the coefficients of the a priori four possible scalar operators are reduced to two and
the tensor operators are forbidden. The phenomenology
of this reduced set of operators in  $B_q\rightarrow l^+l^-$ decays was studied.

The present approach could be extended to other low energy processes
but also combined with EW scale physics to narrow down possible new physics operators.
Finally let us remark that, with the growing experimental data,
the type of correlations discussed here is likely to play an important role in the determination of the nature of the new physics to appear.

\paragraph{Acknowledgments}
We are happy to thank S.~J\"ager for helpful discussions. This work was supported
in part by DOE grants DE-SC000991 and DE-SC0009919. JMC has received funding 
from the People Programme (Marie Curie Actions) of the European Union's 
Seventh Framework Programme (FP7/2007-2013) under REA grant agreement
n PIOF-GA-2012-330458 and acknowledges the Spanish Ministerio de Econom\'ia y 
Competitividad and european FEDER funds under the contract FIS2011-28853-C02-01
for support.

\end{document}